\documentclass[aps,pre,twocolumn,showpacs,superscriptaddress,floatfix]{revtex4}

\usepackage{graphicx}%
\usepackage{dcolumn}
\usepackage{amsmath}
\usepackage{latexsym}

\begin{document}

\def\bea{\begin{eqnarray}}
\def\eea{\end{eqnarray}}
\def\beq{\begin{equation}}
\def\eeq{\end{equation}}
\def\f{\frac}
\def\k{\kappa}
\def\h{\eta}
\def\sx{\sigma_{xx}}
\def\sy{\sigma_{yy}}
\def\sxy{\sigma_{xy}}
\def\e{\epsilon}
\def\ve{\varepsilon}
\def\ex{\epsilon_{xx}}
\def\ey{\epsilon_{yy}}
\def\exy{\epsilon_{xy}}
\def\be{\beta}
\def\D{\Delta}
\def\th{\theta}
\def\r{\rho}
\def\a{\alpha}
\def\s{\sigma}
\def\kb{k_B}
\def\la{\langle}
\def\ra{\rangle}
\def\nn{\nonumber}
\def\bu{{\bf u}}
\def\bn{\bar{n}}
\def\br{{\bf r}}
\def\up{\uparrow}
\def\dn{\downarrow}
\def\S{\Sigma}
\def\dg{\dagger}
\def\d{\delta}
\def\p{\partial}
\def\l{\lambda}
\def\G{\Gamma}
\def\o{\omega}
\def\g{\gamma}
\def\kv{\bar{k}}
\def\ha{\hat{A}}
\def\hv{\hat{V}}
\def\hg{\hat{g}}
\def\hG{\hat{G}}
\def\hTT{\hat{T}}
\def\vv{\vec{v}}
\def\rv{\vec{r}}
\def\hf{\frac{1}{2}}

\title{Anomalous structural and mechanical properties of solids
confined in quasi one dimensional strips} 
\author{Debasish Chaudhuri}
\affiliation{
Max Planck Institute for the Physics of Complex Systems, 
N{\"o}thnitzer Strasse 38, 01187 Dresden, Germany
}
\email{debc@pks.mpg.de}
\author{Surajit Sengupta}
\affiliation{
S. N. Bose National Center for Basic Sciences, JD Block, Sector 3,
Salt Lake, Kolkata - 700098, India
}
\email{surajit@bose.res.in}

\date{\today}
\begin{abstract}
We show using computer simulations and mean field theory that a system
of particles in two dimensions, when confined laterally by a pair of
parallel hard walls within a quasi one dimensional channel,
possesses several anomalous structural and mechanical properties not
observed in the bulk. Depending on the density $\rho$ and the distance 
between the walls $L_y$, the system shows structural characteristics 
analogous to a weakly modulated liquid, a strongly modulated smectic, a 
triangular solid or a buckled phase. At fixed $\rho$, a change in 
$L_y$ leads to many reentrant discontinuous transitions involving changes 
in the number of layers parallel to the confining walls depending 
crucially on the commensurability of inter-layer spacing with $L_y$. 
The solid shows resistance to elongation but not to shear. 
When strained beyond the elastic limit it fails undergoing plastic 
deformation but surprisingly, as the strain is reversed, the material recovers 
completely and returns to its original undeformed state. 
We obtain the phase diagram from mean field theory and finite 
size simulations and discuss the effect of fluctuations.  
\end{abstract}
\maketitle

\section{Introduction}
\label{intro}
 
Recent studies on various confined 
systems~\cite{buckled-1,fortini,peeters,doyle-1,doyle-2,degennes,landman,myfail,teng,peeters2,leiderer,lai,bubeck,peeters3,pieranski-1,ayappa,abhi,hensler} 
have shown the possibility of obtaining many different 
structures and phases depending on the range of interactions and 
commensurability of a microscopic length scale of the system
with the length scale of confinement. 
Since internal structures often play a crucial role in determining
local dynamical properties like asymmetric diffusion, 
viscosity etc. and phase behavior~\cite{doyle-2}, such studies are of 
considerable practical interest. 
Charged particles interacting via a screened Coulomb
potential when confined in a one dimensional parabolic well 
showed many zero temperature layering transitions~\cite{peeters}. 
At high temperatures this classical Wigner crystal 
melts and the melting temperature
shows oscillations as a function of density~\cite{peeters}. Such oscillations
are characteristic of confined systems, arising out of commensurability.
In Ref.~\cite{degennes,landman} a similar layering transition is found
in which the number of layers in a confined liquid with smectic modulations
changes discretely as the wall-to-wall separation is increased. 
The force between the walls oscillates as a function of this separation
depending crucially on the ratio of the separation to the thickness of the 
smectic layers. 
Confined crystals always align one of the lattice planes along the direction
of confinement~\cite{peeters,doyle-1} and confining walls generate 
elongational asymmetry in the local density profile along the walls even 
for the slightest incommensuration~\cite{myfail}. 
For long ranged interactions, extreme
localization of wall particles has been observed~\cite{peeters2,doyle-1}. 
Hard spheres confined within a two dimensional slit bounded
by two parallel hard plates show a rich phase 
behavior~\cite{pieranski-q2d,Neser,buckled-1,buckled-2,buckled-3,fortini}. 
For separations of one to five hard sphere diameters a phase diagram 
consisting of a dazzling array of up to $26$ distinct crystal structures 
has been obtained~\cite{fortini,pieranski-q2d}. Similarly confined 
systems with more general interactions also show a similar array of 
phases\cite{ayappa}.
In an earlier experiment on confined steel balls in quasi one dimensions (Q1D),
i.e. a two dimensional system of particles confined in a narrow channel by 
parallel walls ({\em lines}), vibrated to
simulate the effect of temperature, layering transitions, phase coexistence
and melting was observed~\cite{pieranski-1}.
A recent study on trapped Q1D solid in contact of its own liquid 
showed layering transitions with increase in trapping potential\cite{abhi}.
Layering fluctuations in such a solid has crucial impact on the 
sound and heat transport across the system\cite{dcacss,acdcss}.

In this paper, we show that a Q1D solid strip of length $L_x$ confined 
within parallel, hard, one dimensional walls separated by a distance $L_y$ 
has rather anomalous properties. These are quite 
different from bulk systems in one, two or three dimensions as well as 
from Q1D solid strips with periodic boundary conditions (PBCs). We list 
below the main characteristics of a Q1D confined strip that we demonstrate
and discuss in this paper:

\begin{enumerate}
\item{\em Re-entrant layer transitions:~}The nature of the system in 
Q1D depends crucially on the density $\rho$ and width of the channel 
$L_y$. The number of layers of particles depends on the ratio of the 
interlayer spacing (fixed mainly by $\rho$) and $L_y$.   
With increase in channel width $L_y$ at a fixed $\rho$ we find many re-entrant
discontinuous transitions involving changes in the number of layers parallel 
to the confining direction. The {\em phase diagram} in the $\rho- L_y$ 
plane is
calculated from Monte Carlo simulations of systems with finite size as well 
as mean field theory (MFT). While all the phases show density modulations
in the direction perpendicular to the wall, we identify distinct analouges 
of the bulk phases viz. modulated liquid, smectic, triangular solid and 
buckled solid.

\item {\em Anomalous elastic moduli:~} A solid
characterized by a periodic arrangement of particles offers resistance to 
elongation as well as shear. The Q1D confined solid is shown to have a 
large Young's modulus which offers resistance to tensile deformations. 
On the other hand the shear modulus of the system is vanishingly small so 
that layers of the solid parallel to the confinining wall may slide past 
each other without resistance. 

\item {\em Reversible failure:~} Under externally imposed tensile strain the 
deviatoric stress shows an initial linear rise up to a limiting
value which depends on $L_y$. On further extension the stress 
rapidly falls to zero accompanied by a reduction in the number of solid 
layers parallel to the hard walls by one. However, this failure is reversible 
and the system {\em completely recovers the initial structure once the strain 
is reduced} quite unlike a similar solid strip in presence of PBCs in both the
directions. The critical strain for failure by this mechanism
decreases with increasing $L_y$ so that thinner strips
are {\em more} resistant to failure. We show that this reversibility is
related to the anomalies mentioned above. Namely, the confined solid,
though possessing {\em local} crystalline order, retains the ability to flow and
regenerate itself. In this manner portions of the Q1D confined solid behave
like coalescing liquid droplets. A preliminary study of this reversible failure
mechanism was reported in Ref.\cite{myfail}.

\item {\em Displacement fluctuation and solid order:~} Long wavelength 
displacement fluctuations in Q1D are expected to destabilize crystalline order 
beyond a certain length scale. While we always observe this predicted growth of 
fuctuations, in the case of confined system, the {\em amplitude} depends 
crucially on the wall separation. If $L_y$ is incommensurate with the 
interlayer spacing, then local crystalline order is destabilized. Otherwise,  
fluctuations are kinetically suppressed in the confined system at high 
densities. 
Finite size effects also tend to saturate the growth of fluctuations.
Solid strips of finite length therefore  
exhibit apparent crystalline order at high densities both in simulations as 
well as in experiments\cite{peeters}. 
\end{enumerate}

We have used an idealized model 
solid to illustrate these phenomena. Our model solid has particles (disks)
which interact among themselves only through excluded volume or ``hard''
repulsion.  We have reasons to believe, however, that for the questions 
dealt with in this paper, the detailed nature of the inter particle 
interactions are relatively irrelevant and system behavior is largely 
determined by the nature of confinement and the constraints. 
Our results may be directly verified in experiments on sterically stabilized
``hard sphere'' colloids\cite{colbook} confined in glass channels
and may also be relevant for similarly confined atomic systems interacting
with more complex potentials. Our results should hold, at least qualitatively, 
for systems with fairly steep repulsive 
interactions\cite{acdcss,ricci,ricci2}.

This paper is organized as follows. In the next section, we  introduce
the model confined solid and discuss the geometry and basic definitions of 
various structural and thermodynamic parameters. We then introduce
the various possible structures with their basic characteristics
in section~\ref{phases}. In section~\ref{results}, this will be followed by 
the results of computer simulations,
in the constant NAT (number, area, temperature) ensemble, exploring the 
deformation and failure properties of this system and the relation of the 
various structures described in section~\ref{phases} to one another.  
In section~\ref{phase-diagram}, we  provide a 
finite size phase diagram obtained from simulations and compare it with an 
MFT calculation. In section~\ref{conclusion} we discuss our results
with emphasis on the role of long wave length fluctuations in the
destruction of crystalline order in low dimensions and
conclude giving some outlook for future work.

\section{The model and method}
\label{system}
The bulk system of hard disks where particles $i$ and $j$, in two dimensions, 
interact with the potential $V_{ij} = 0$ for $|{\bf r}_{ij}| > {d}$ and 
$V_{ij} = \infty$ for $|{\bf r}_{ij}| \leq {d}$, with ${d}$ 
the hard disk diameter and ${\bf r}_{ij} = {\bf r}_j - {\bf r}_i$ the 
relative position vector 
of the particles, has been extensively\cite{al,zo,web,jaster,sura-hdmelt} 
studied. Apart from being easily 
accessible to theoretical treatment\cite{hansen-macdonald}, experimental systems
with nearly ``hard'' interactions\cite{colbook} are 
available. The hard disk free energy is entirely entropic in 
origin and the only thermodynamically relevant variable is the number density   
$\rho = N/A$ or the packing fraction $\eta = (\pi/4) \rho {d}^2$. 
Accurate computer simulations\cite{jaster} of hard 
disks show that for $\eta > \eta_f = 0.719$ the system exists as a triangular 
lattice which melts below $\eta_m = 0.706$. The melting occurs possibly
through a two step continuous transition from solid to liquid via an 
intervening hexatic phase\cite{jaster,sura-hdmelt}. 
Elastic constants of bulk hard disks have been 
calculated in simulations\cite{branka, sura-hdmelt}.
The surface free energy of the hard disk system in contact with a hard wall
has also been obtained\cite{hartmut} taking care that the 
dimensions of the system are compatible with a strain-free 
triangular lattice. 

\begin{figure}[t]
\begin{center}
\includegraphics[width=8.0cm]{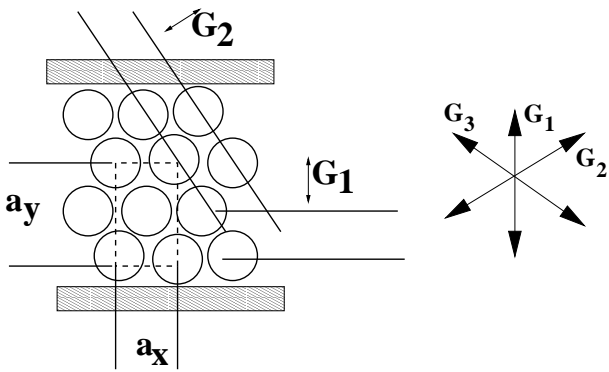}
\end{center}
\caption{The confined solid is shown along with the centered rectangular (CR) unit cell. For 
an unstrained triangular lattice $a_x = a_0$ and $a_y = \sqrt{3} a_0$. ${\bf
 G_1}$,  ${\bf G_2}$ and ${\bf G_3}$ denote the directions of the three reciprocal lattice vectors (RLV). Note that the third reciprocal lattice direction ${\bf G_3}$ is equivalent to ${\bf G_2}$, even in the presence of the walls.
}
\label{wallpic}
\end{figure}

The wall- particle interaction potential $V_{\rm wall}(y) = 0$ for 
$ {d/2} < y < L_y - {d/2}$ and $ = \infty$ otherwise. Here, evidently, $L_y$ is
the width of the channel. The length of the channel is $L_x$ with 
$L_x \gg L_y$. Periodic boundary conditions are assumed in the $x$ 
direction(Fig.\ref{wallpic}). 

Before we go on to describe the various 
phases we observe in this model system, it is instructive to consider how 
a triangular lattice (the ground state configuration) may be accomodated 
within a pair of straight hard walls. 
For the channel to accommodate $n_{l}$ layers of a homogeneous, triangular 
lattice with lattice parameter $a_0$ of hard disks of diameter ${d}$, 
(Fig.\ref{order}) it is required that, 
\begin{equation}
L_y = \frac{\sqrt{3}}{2}(n_{l} - 1) a_0 + {d} ~.
\label{perfect}
\end{equation}
For a system of constant number of particles and $L_y$, $a_0$ is a function of
packing fraction $\eta$ alone. 
We define $\chi(\eta, L_y) = 1 + 2(L_y - {d})/\sqrt{3} a_0$, so that 
the above condition reads $\chi = {\rm integer} = n_{l}$ 
(the {\em commensurate} configuration) and violation
of Eq.(\ref{perfect}) implies a rectangular strain away from the reference
triangular lattice of $n_l$ layers. The lattice parameters of a centered 
rectangular (CR) unit cell are $a_x$ and $a_y$ (Fig. \ref{wallpic}). 
In general, for a CR lattice with a given  $L_y$, 
$a_y = 2 (L_y - {d})/(n_l-1)$ and
$a_x = 2/\rho a_y$, ignoring vacancies. There are two distinct classes of 
close packed  planes in the CR lattice. 
Due to the presence of confinement, even
for a triangular lattice, the set of planes with reciprocal lattice vector
(RLV) ${\bf G_1} = \hat{y} \frac{4\pi}{a_y}$ perpendicular to the walls
are distinct from the equivalent set of planes with the RLV's
${\bf G_2} = \hat{x} \frac{4\pi}{a_y}\cos(\frac{\pi}{6}) + \hat{y} \frac{4\pi}{a_y}\sin(\frac{\pi}{6})$ and 
${\bf G_3} = \hat{x} \frac{4\pi}{a_y}\cos(\frac{\pi}{6}) - \hat{y} \frac{4\pi}{a_y}\sin(\frac{\pi}{6})$ (Fig.\ref{wallpic}).

\begin{figure}[t]
\begin{center}
\includegraphics[width=4.0cm]{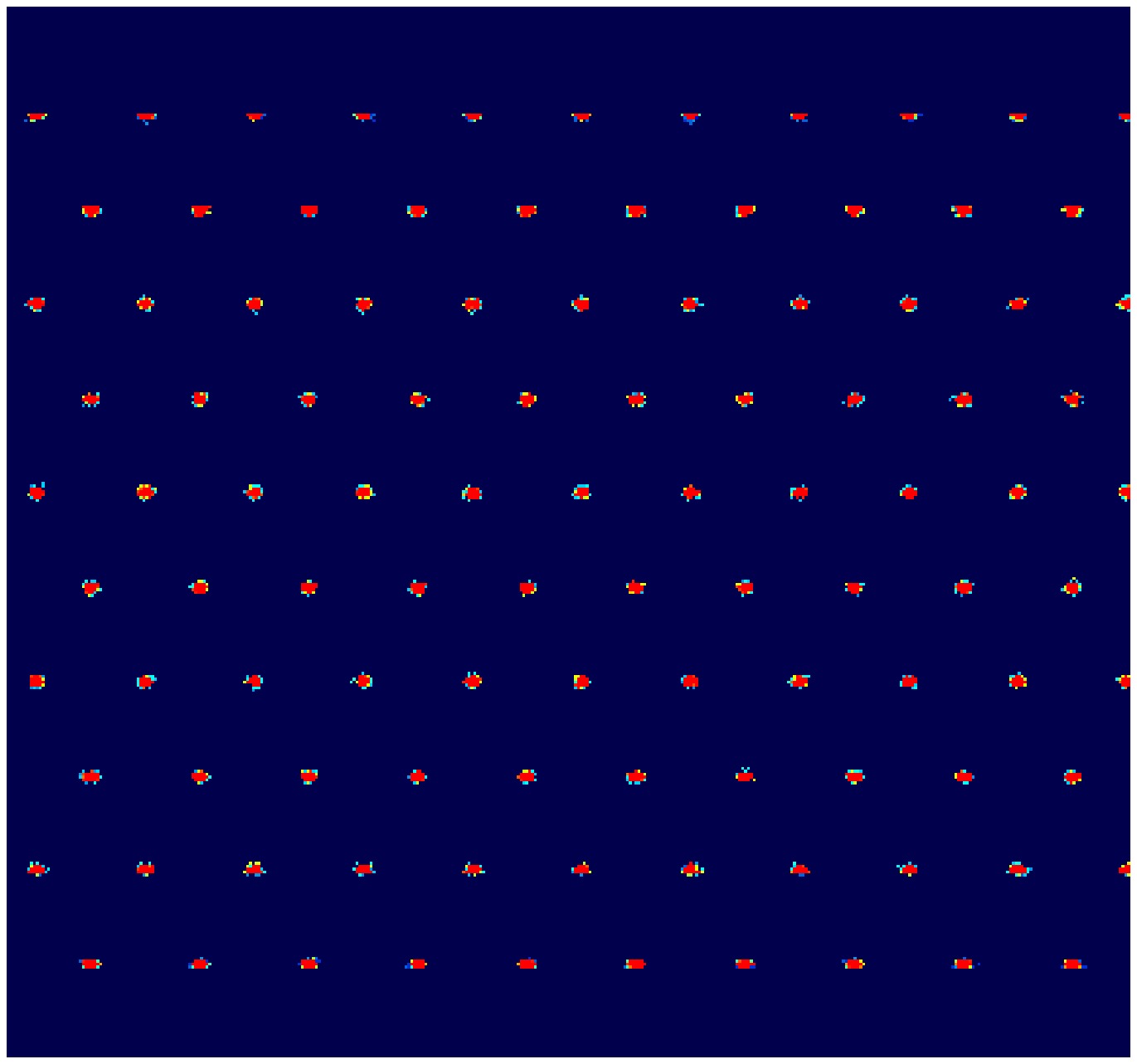}
\includegraphics[width=4.0cm]{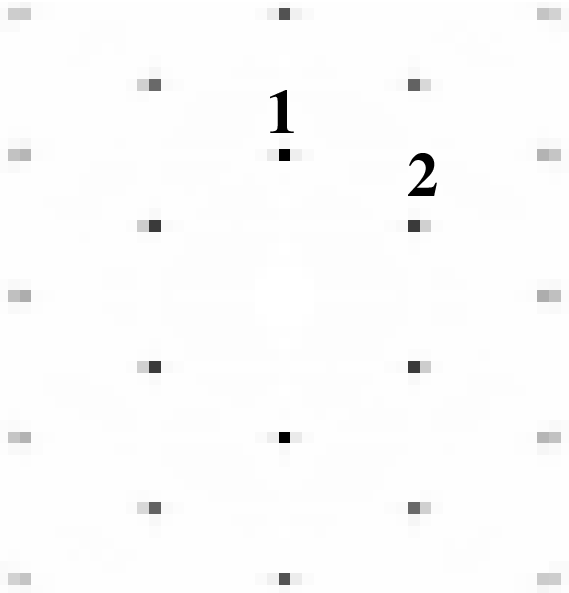}
\end{center}
\caption{(Colour online)
Solid: Left panel shows a picture of $10^3$ uncorrelated, superimposed 
configurations of a high density ($\eta=0.85$) solid phase. The wall to wall 
separation 
is commensurate with a ten layered solid at the given density. 
The colour code is such that red (light) 
means high local density and blue (dark) means low density. 
The right panel shows the 
corresponding structure factor which shows a pattern typical for a 
two dimensional triangular solid.  
}
\label{sld}
\end{figure}

Anticipating some of the discussion in section \ref{results}, we 
point out two different but equivalent ``pictures'' for studying the 
deformation behavior of confined narrow crystalline strips. In the first 
picture, the stress is regarded as a function of the ``external'' strain.  
Using the initial triangular solid (packing fraction $\eta_0$) 
as reference, the external strain associated with changing $L_x$ 
at a constant $N$ and $L_y$
is $\epsilon =(L_x-L_x^0)/L_x^0$.
In this case we obtain the stress as an oscillating non-monotonic function
of $\epsilon$. On the other 
hand, internally, the solid is, free to 
adjust $n_l$ to decrease its energy (strain). Therefore, one may, 
equivalently, calculate
strains with respect to a reference, distortion-free, triangular lattice 
at $\eta$. Using the definition  $\varepsilon_d = 
\varepsilon_{xx} - \varepsilon_{yy} = (a_x - a_0)/a_0 - (a_y - \sqrt{3}a_0)/\sqrt{3}a_0 = a_x/a_0 - a_y/\sqrt{3}a_0$ and the expressions for 
$a_x$, $a_y$ and  $a_0 = 2(L_y-d)/\sqrt{3}(\chi -1) $  we obtain, 
\begin{equation}
\varepsilon_d =  \frac{n_l - 1}{\chi - 1} - \frac{\chi - 1}{n_l - 1},
\label{strain}
\end{equation}
where the number of layers $n_l$ is the nearest integer to $\chi$ so that 
$\varepsilon_d$ has a discontinuity at half~-integral values of $\chi$. 
For large $L_y$ this discontinuity and $\varepsilon_d$ itself vanishes as 
$1/L_y$ for all $\eta$. This ``internal'' strain $\varepsilon_d$ is related 
non-linearly to $\epsilon$ and may remain small even if $\epsilon$ is large.
The stress is always a monotonic function of $\varepsilon_d$. 
Note that a pair of variables $\eta$ and $L_y$ (or alternately 
$\epsilon$ and $\chi$) uniquely fixes the state of the system. 

\begin{figure}[t]
\begin{center}
\includegraphics[width=8.6cm]{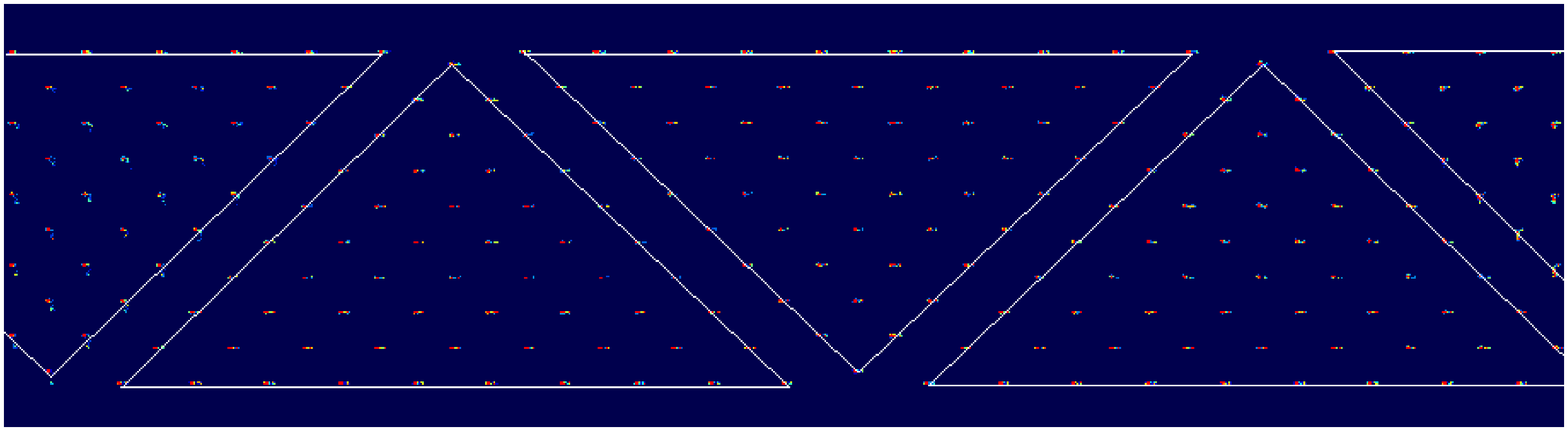}
\includegraphics[width=4.0cm]{buckl-sk-real.eps}
\end{center}
\caption{(Colour online)
Buckled phase: A small incommensuration is introduced by increasing the 
wall to wall separation from the value commensurate to a high density 
ten layered triangular solid at $\eta=0.89$. This reduces the packing fraction
to $\eta=0.85$ and produces this phase. 
The upper panel shows a picture of  $10^3$ superimposed 
configurations. The colour code for the local densities is the same as 
Fig.\ref{sld}. Note that different portions of triangular solid are
displaced along $y$- direction to span the extra space introduced between the
walls. Lines are drawn to identify this shift in triangular regions.
The lower panel shows the corresponding structure factor where
the peak in ${\bf G}_1$ direction is diminished. Some extra weak peaks 
corresponding to super-lattice reflections appear at lower values of the 
wave-number.
}
\label{bkld}
\end{figure}

We have carried out extensive Monte Carlo (MC) simulations in the
constant NAT ensemble using standard Metropolis updates for hard disks 
(i.e. moves are rejected if they lead to overlaps). The initial condition
for our simulations is the perfectly commensurate triangular lattice with
a fixed number of layers $n_l$ and at a fixed packing fraction $\h$. After
equilibration for typically about $10^6$ Monte Carlo steps (MCS), a strain 
$\e$ is imposed by 
rescaling $L_x$. Since the walls
are kept fixed, this strain reduces $\h$. The sequence of phases and the 
nature of the transitions among them induced by this strain is discussed
below.

\section{Structures and Phases}
\label{phases}

In Q1D long wavelength fluctuations\cite{ricci2} are expected to destroy 
all possible order except for those imposed by {\em explicit} 
(as opposed to spontaneous) breaking of symmetry.
The confining potential, in the case of a Q1D system in a hard 
wall channel, explicitly breaks the continuous rotational
symmetry down to a ${\cal Z}_2$ symmetry (rotations by angle $\pm \pi$). This 
immediately leads to all $2n$-atic bond orientational orders, like
nematic ($n = 1$), square ($n = 2$), hexatic ($n = 3$) etc. which remain
nonzero throughout the phase diagram. 
This situation is therefore similar to a system in a periodic 
laser potential\cite{frey-lif-prl,lif-hd,mylif}. 
Apart from orientational symmetry, the confining potential explicitly
breaks translational symmetry perpendicular to the walls leading to a
density modulation in that direction. Any 
fluctuation which leads to global changes in the direction of layering or in 
the layer spacing for fixed wall to wall separation is strongly 
suppressed. 

For finite systems (and over finite - but very long - times)
of confined Q1D strips one obtains  long-lived, metastable ordered phases 
which are observable even in experiments\cite{pieranski-1}. 
It is these  `phases' that we describe in this section. 
For convenience in nomenclature, we continue to use the terms
solid, modulated liquid and smectic to denote these phases, 
though keeping in mind 
that the distinctions are largely quantitative and not qualitative. 
For example,  a weak solid-like local hexagonal modulation is 
present on top of the smectic layering order in what we 
call the smectic phase. The smectic develops continuously and smoothly 
from the modulated liquid. We also denote the sharp changes in observables 
(eg. strength of diffraction peaks) as ``phase transitions'' though they can 
never be classified as true phase transitions in the sense of equilibrium 
thermodynamics. We show, however, that these phase transitions may be 
described within MFTs (section \ref{phase-diagram}) although they might loose 
meaning when fluctuations are considered  in the limit of infinite observation 
time. 

In our description of the phases we make use of the structure factor 
\bea
S_{\bf G} = \left< \frac{1}{N^2} \sum_{j,k = 1}^N 
\exp(-i {\bf G}.{\bf r}_{jk})\right> ,\nn
\eea
where ${\bf r}_{ij} = {\bf r}_j - {\bf r}_i$ with ${\bf r}_i$ the position 
vector of particle $i$. We shall use particularly the values of $S_{\bf G}$ 
for the reciprocal lattice vectors ${\bf G} = \pm {\bf G_1}(\eta)$ and 
${\bf \pm G_2}(\eta)$. Notice that ${\bf \pm G_2}(\eta)$ and 
${\bf \pm G_3}(\eta)$ as shown in Fig.\ref{wallpic} 
are equivalent directions. A plot of $S_{\bf G}$ for ${\bf G} = (G_x,G_y)$
in the two dimensional plane gives the diffraction pattern observable in 
scattering experiments. For every phase discussed below, the diffraction 
pattern {\em always} shows at least two peaks corresponding to $\pm {\bf G}_1$
since they represent the density modulation imposed by the walls. The 
relative strengths of these peaks, of course, depend on the structure of the 
phase concerned.    
 
If the separation between the hard walls is kept commensurate such 
that $\chi=n_l$, an integer, at high density we obtain a perfect 
two dimensional {\em triangular solid} (Fig.\ref{sld}). 
The solid shows a diffraction pattern which is typical for a two dimensional
triangular crystal. We show later that appearances can be deceptive, however.
This triangular ``solid'' is shown to have zero shear modulus which would mean 
that it can flow without resistance along the length of the channel like a 
liquid. Stretching the solid strip lengthwise, on the other hand, costs energy 
and is resisted. The strength of the diffraction peaks decreases rapidly with 
the order of the diffraction. In strictly two dimensions this is governed by
a non-universal exponent dependent on the elastic constants \cite{kthny1}. 
In Q1D this decay should be faster. However, larger system sizes 
and averaging over a large number of configurations would be required to 
observe this decay, since constraints placed by the hard walls make the system 
slow to equilibrate at high densities. For a general value of $\chi$ the 
lattice is strained which shows up in the relative intensities of the 
peaks in the diffraction pattern corresponding to ${\bf G}_2$ and ${\bf G}_3$.   

\begin{figure}[t]
\begin{center}
\includegraphics[width=8.0cm]{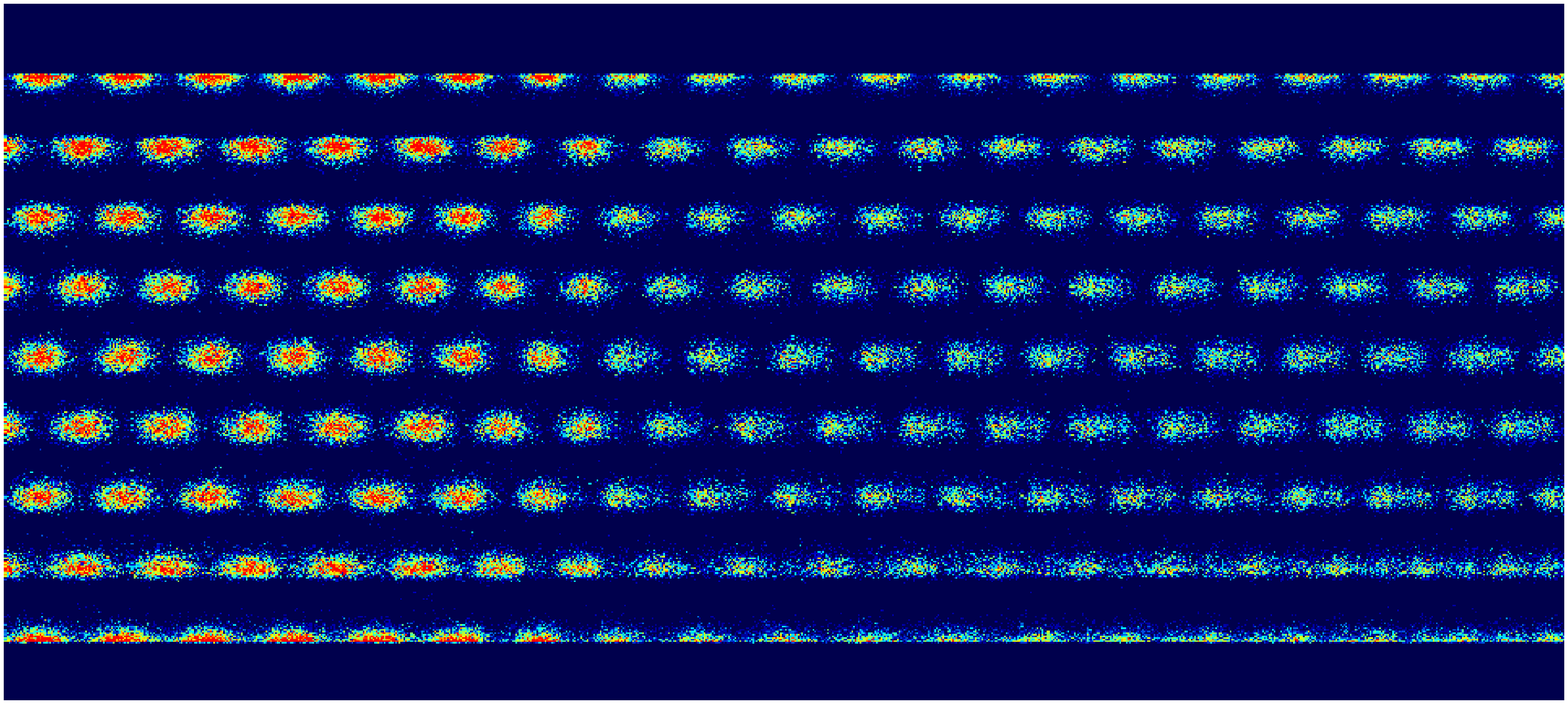}
\vskip .1cm
\includegraphics[width=4.0cm]{smec-73-etay.eps}
\includegraphics[width=4.0cm,angle=90]{Sk-smectic-73.eps}
\end{center}
\caption{(Colour online) Smectic: 
A confined triangular solid at $\eta=0.85$ is strained in $x$-direction
to a packing fraction $\eta=0.73$.
The upper panel shows a picture of $10^3$ superimposed 
configurations. The colour code for local density is same as Fig.\ref{sld}.
The lower left panel shows the 
density modulation in $y$- direction.
The lower right panel shows the corresponding structure factor where the
only remaining strong peaks are in ${\bf G}_1$ direction identifying a 
smectic order (solid-like in $y$-direction and liquid-like in other 
directions).
This smectic phase, however, possesses hexagonal modulations 
producing weak triangular order leading to faint secondary spots in 
the structure factor.
}
\label{smec}
\end{figure}

\begin{figure}[t]
\begin{center}
\includegraphics[width=8.0cm]{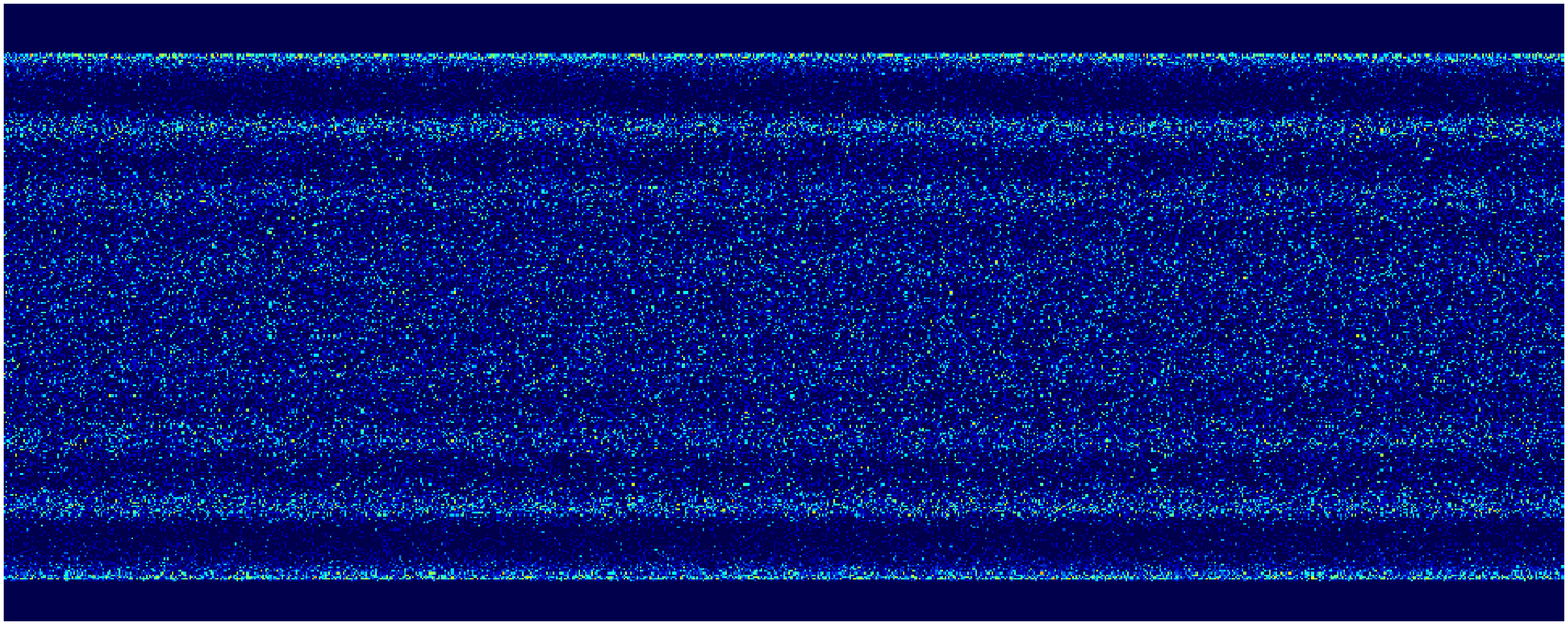}
\vskip .1cm
\includegraphics[width=4.0cm]{modliq-60-etay.eps}
\includegraphics[width=4.0cm]{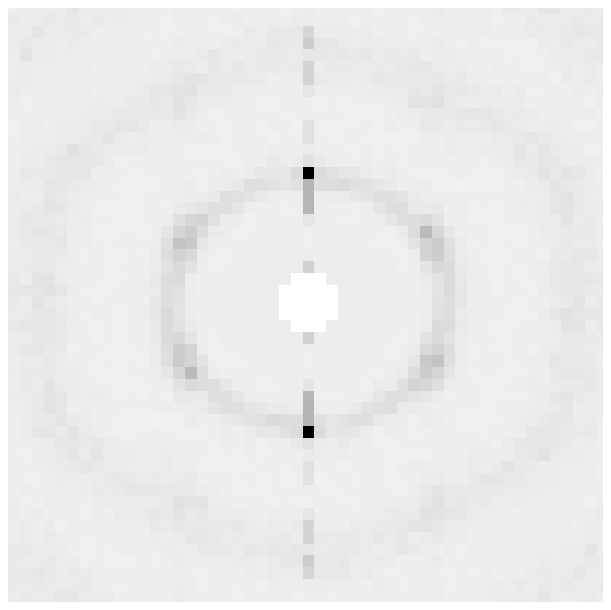}
\end{center}
\caption{(Colour online)
Modulated liquid: 
A confined triangular solid at $\eta=0.85$ is strained in $x$-directon
reducing the density to $\eta=0.6$.
The upper panel shows a picture of $10^3$ superimposed 
configurations. The colour code of local densities is the same 
as Fig.\ref{sld}. The lower left panel shows the density 
modulation in $y$- direction which is like the smectic phase but the 
modulation dies out at the center. The structure factor in the lower right  
panel shows a ring like pattern which is a typical signature of liquid. 
Superimposed on this are strong smectic-like peaks in ${\bf G}_1$ direction. 
}
\label{lqd}
\end{figure}
A little extra space introduced between the walls starting from a high density
solid phase gives rise to buckling instability in $y$- direction and the system
breaks into several triangular solid regions along the $x$- direction 
(Fig.\ref{bkld}).
Each of these regions slide along close packed planes (corresponding to 
the reciprocal directions ${\bf G}_2$ or ${\bf G}_3$) with respect to one 
another. This produces a buckling wave with displacement in $y$- direction 
travelling along the length of the solid. 
In conformity with the quasi two dimensional analog \cite{buckled-1, 
buckled-2, buckled-3} we call this the {\em buckled solid} and it interpolates
continuously from $\chi = n_l$ to $ n_l\pm1$ layers. This phase can also 
occur due to the introduction of a compressional strain in $x$-direction 
keeping $L_y$ fixed. The diffraction pattern 
shows a considerable weakening of the spots corresponding to planes parallel 
to the walls $S_{\bf G_1}$ 
together with generation of extra spots at smaller wave-number corresponding 
to the buckled super-lattice. The diffraction pattern is therefore almost 
complementary to that of the smectic phase to be discussed below.
We do not observe the buckled solid at low densities
close to the freezing transition. Large fluctuations at such densities 
lead to creation of bands of the smectic phase within a solid eventually 
causing the solid to melt.

At low enough densities or high enough in-commensuration ($\chi$ half integral)
the elongated density profiles in the lattice planes parallel to the walls can 
overlap to give rise to 
what we denote as the {\em smectic} phase (Fig.\ref{smec}) in which
local density peaks are smeared out in $x$- direction but are clearly
separated in $y$-direction giving rise to a layered structure. The diffraction
pattern shows only two strong spots ($S_{\bf G_1}$) which is typical 
corresponding to the  symmetry of a smectic phase. We use this fact as the 
defining principle for this phase.
Note that unlike usual smectics there is no orientational ordering of
director vectors of the individual particles since hard disks are isotropic.

At further lower densities the relative displacement fluctuations between 
neighbors diverges and the diffraction pattern shows a ring- 
like feature typical of a liquid which appears together with the smectic like 
peaks in the direction perpendicular to the walls. We monitor this using  
the relative Lindemann parameter\cite{pcmp}, 
given by,
\bea
l =\la ({u^x}_i - {u^x}_j)^2\ra /a_x^2 + 
\la ({u^y}_i - {u^y}_j)^2\ra/a_y^2
\eea
where the angular brackets denote averages over configurations, 
$i$ and $j$ are nearest neighbors and ${u^{\alpha}}_i$ is the $\alpha$-th 
component of the displacement of particle $i$ from it's mean position. 
This phase is a {\em modulated liquid} (Fig.\ref{lqd}). 
The density modulation decays away 
from the walls and for large $L_y$, the density profile in the middle of the 
channel becomes uniform. A clear structural difference between smectic and 
modulated liquid is the presence of the ring pattern in the structure factor 
of the latter, a characteristic of liquid phase (compare Fig.\ref{smec} and 
\ref{lqd}).
We must emphasize here again that the distinction between a 
modulated liquid and a smectic is mainly a question of degree
of the layering modulation. The two structures merge continuously into one 
another as the density is increased. Also when the modulated liquid 
co-exists with the solid, the layering in the former is particularly 
strong due to proximity effects.

\section{Mechanical Properties and Failure}
\label{results}

In this section we present the results for the mechanical behavior 
and failure of the Q1D confined solid under tension. We know that   
fracture in bulk solids occurs by the nucleation and growth of 
cracks\cite{griffith,marder-1,marder-2,langer}. The interaction of 
dislocations or zones of plastic deformation\cite{langer,loefsted} with 
the growing crack tip determines the failure mechanism viz. either ductile 
or brittle fracture. Studies of the fracture of single-walled  
carbon nanotubes\cite{SWCNT-1,SWCNT-2} show failure driven by bond-breaking  
which produces nano cracks which run along the tube circumference leading to
brittle fracture. Thin nano-wires of Ni are 
known\cite{nano-wire-1,nano-wire-2} to show ductile failure with 
extensive plastic flow and amorphization. 
We show that the Q1D confined solid behaves anomalously, quite unlike any of 
the mechanisms mentioned above. It shows reversible plastic deformation and 
failure in the constant extension ensemble. The failure occurs by the 
nucleation and growth of smectic regions which occur as distinct bands 
spanning the width of the solid strip. 
\begin{figure}[t]
\begin{center}
\includegraphics[width=8.6cm]{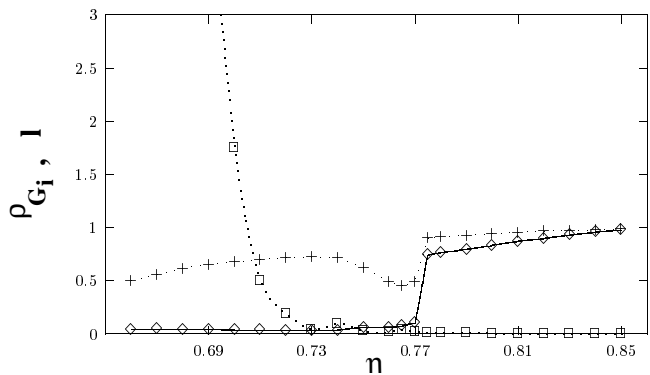}
\end{center}
\caption{Results of NAT ensemble MC simulations of 
$N = n_x \times n_y = 65 \times 10$ hard disks confined between two 
parallel hard walls separated by a distance
$L_y = 9\,{d}$. For each $\eta$, the 
system was allowed to run for $10^6$ MCS  and data 
averaged over a further $10^6$ MCS. 
At $\eta = 0.85$ we have a strain free triangular lattice.
Plots show the structure factors $S_{\bf G_i}, i = 1 (+),2(\diamond)$
for RLVs ${\bf G_i}(\eta)$, averaged over 
symmetry related directions, as a function of $\eta$. 
Also plotted in the same graph is the Lindemann parameter $l(\Box)$.
The lines in the figure are a guide to the eye.} 
\label{order}
\end{figure}

We study the effects of strain on the hard disk triangular solid for a 
$L_y$ large enough to accommodate only a small number of layers 
$n_l \sim 9 - 25$. 
The Lindemann parameter $l$ diverges at the melting transition \cite{Zahn}. We 
also compute the quantities $S_{{\bf G = G}_1}$ and $S_{{\bf G = G}_2}$.

\begin{figure}[t]
\begin{center}
\includegraphics[width=8.6cm]{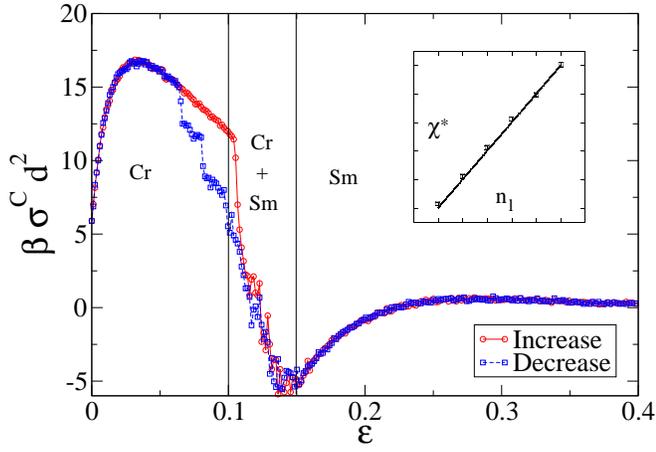}
\end{center}
\caption{(Colour online)
A plot of the deviatoric stress $\sigma^C$ versus external strain 
$\epsilon$ obtained from our MC simulations of $65 \times 10$
hard disks initially at $\eta = 0.85$ and $L_y=9d$. Data 
is obtained by 
holding at each strain value for $2\times10^4$ MCS and averaging 
over a further $3\times10^4$ MCS.  The entire  cycle of increasing 
$\epsilon (\circ)$ and decreasing to zero $(\Box)$ 
using typical parameters appropriate for an atomic system,
corresponds to a real frequency of $\omega\approx 25 {\rm K\,Hz}$. 
Results do not essentially change for $\omega=1 {\rm K\,Hz}\,- 1{\rm M\,Hz}$.
The line is a guide to the eye. The vertical lines mark the limit of 
stability of the crystal (Cr), the two phase region (Cr+Sm) and the onset
of the smectic phase (Sm).
Inset shows the 
variation of the critical $\chi^{\ast}$ with 
$n_l$, points: simulation data; line: $\chi^{\ast} = n_l-1/2$.}
\label{stress}
\end{figure}

In Fig.\ref{order} we show how $S_{\bf G_2},~  S_{\bf G_1}$ and
$l$ vary as a function of externally imposed elongational strain that
reduces $\eta$.
Throughout, $S_{\bf G_2} <  S_{\bf G_1} \neq 0$. This is a 
consequence of the hard wall constraint\cite{hartmut} which induces
an oblate anisotropy in the local density peaks of the solid off from
commensuration (nonintegral $\chi$). 
As $\eta$ is decreased both $S_{\bf G_1}$ and $S_{\bf G_2}$ show 
a sharp drop at $\eta = \eta_{c_1}$ 
where $\chi = \chi^{\ast}\approx n_l - 1/2$
[Fig. \ref{stress} (inset)~]. 
For $\eta < \eta_{c_1}$ we get  
$S_{\bf G_2} = 0$ with $S_{\bf G_1} \not= 0$ signifying 
a transition from crystalline to smectic like  order. 
The Lindemann parameter $l$ remains zero and diverges only below 
$\eta = \eta_{c_3}(\approx \eta_m)$ indicating a finite-size-
broadened ``melting'' of the smectic to a modulated liquid phase. 

To understand the mechanical response of 
the confined strips, we compute the deviatoric stress built up in the system
as a function of applied strain. The stress tensor of a bulk system
interacting via two body central potential has two parts: (i)
A kinetic part $\sigma^K_{\lambda\nu}$ leading to an isotropic pressure 
and (ii) A virial term due to inter-particle interaction 
$\sigma^{int}_{\lambda\nu}$. 
The free-particle-like
kinetic component of the stress $\be \sigma^K_{\lambda\nu}=-\r \d_{\l\nu}$
and the component due to inter-particle interaction
$\sigma^{int}_{\lambda\nu}=-\la \sum_{<ij>} r_{ij}^\l f_{ij}^\nu\ra/S$ 
with $f^\nu_{ij}$ the $\nu$-th component of inter-particle force, 
$S=L_x L_y$ area of the strip. 
The expression for the stress tensor for the bulk system of hard disks 
translates to\cite{elast}
\bea
\be \s_{\l\nu} d^2 = -\frac{d^2}{S}\left(\sum_{<ij>}\left< 
    \delta(r_{ij}-d)~\frac{r^\lambda_{ij}r^\nu_{ij}}{r_{ij}}\right> +
  N\delta_{\lambda\nu} \right).
\eea 

The presence of walls gives rise to a potential which varies only in
the $y$-direction perpendicular to the walls. Therefore, strains $\ex$ and
$\e_{xy}$ do not lead to any change in the wall induced potential. As a
consequence the conjugate stresses for the confined system 
$\s^C_{xx}=\s_{xx}$ and $\s^C_{xy}=\s_{xy}$. However, a strain $\ey$ does lead
to a change in potential due to the walls and therefore a new term in the 
expression for conjugate stress appears\cite{mylif-large},
 $\sy^C=\s_{yy}+\s^w$ with
$\s^w=-\la \sum_{<iw>} f^w_i y_{iw}\ra/S$ where
$w$ denotes the two confining walls. This expression can be easily understood
by regarding the two walls as two additional particles of infinite
mass\cite{varnik}. 
Thus, to obtain the component of the total 
stress normal to the walls from MC simulations we use
\bea
\be\sy^C d^2 &=& \be\sy d^2 
- \f{d^2}{S}\left[ \left\la \sum_i y_i \d(y_i-d/2)\right\ra \right. \nn\\
&+&\left.  \left\la \sum_i (L_y-y_i) \d(L_y-y_i-d/2)\right\ra\right].
\eea

The deviatoric stress, $\sigma^C = \sigma^C_{xx} - \sigma^C_{yy}$, 
versus strain, $\epsilon=\ex-\ey$ ($\ey=0$) curve  for confined hard disks is 
shown in Fig. \ref{stress}. For $\eta = \eta_0$ ($\epsilon = 0$) the stress 
due to the inter-particle interaction is 
purely hydrostatic with $\sigma_{xx} = \sigma_{yy}$ as expected; however,
due to the excess pressure from the walls the solid is actually always 
under compression along the $y$ direction, thus $\sx^C>\sy^C$.
At this point the system is perfectly commensurate with channel
width and the local density profiles are circularly symmetric.

Initially, the stress increases linearly, flattening out at the  
onset of plastic behavior at $\eta \stackrel{<}{\sim} \eta_{c_1}$. 
At $\eta_{c_1}$, with the nucleation of smectic bands,
$\,\,\sigma^C$ decreases and eventually becomes negative. 
At $\eta_{c_2}$ the smectic phase spans the entire system and $\sigma^C$ is 
minimum. On further decrease in $\eta$ towards $\eta_{c_3}$,  
$\sigma^C$ approaches $0$ from below (Fig. \ref{stress}) thus forming a loop. 
Eventually it shows a small overshoot, which ultimately goes to
zero, from above, as the smectic smoothly goes over to a more symmetric liquid
like phase -- thus recovering the Pascal's law at low enough densities.
If the strain is reversed by increasing $\eta$
back to $\eta_0$ the entire stress-strain curve is traced back 
with no remnant stress at $\eta = \eta_0$ showing that the 
plastic region is reversible. For the system shown in Fig.\ref{wallpic} 
we obtained $\eta_{c_1} \approx 0.77$,   
$\eta_{c_2} \approx 0.74$ and $\eta_{c_3} \approx 0.7$.
As $L_y$ is increased,
$\eta_{c_1}$ merges with $\eta_{c_3}$ for $n_l \stackrel{>}{\sim} 25$. 
If instead, $L_x$ and $L_y$ are both rescaled to keep $\chi = n_l$ fixed or 
PBCs are imposed in both $x$ and $y$ directions, the 
transitions in the various quantities occur approximately simultaneously
as expected in the bulk system. Varying $n_x$ in the range $50 - 5000$ produces
no qualitative change in most of the results.

\begin{figure}[t]
\begin{center}
\includegraphics[width=8.6cm]{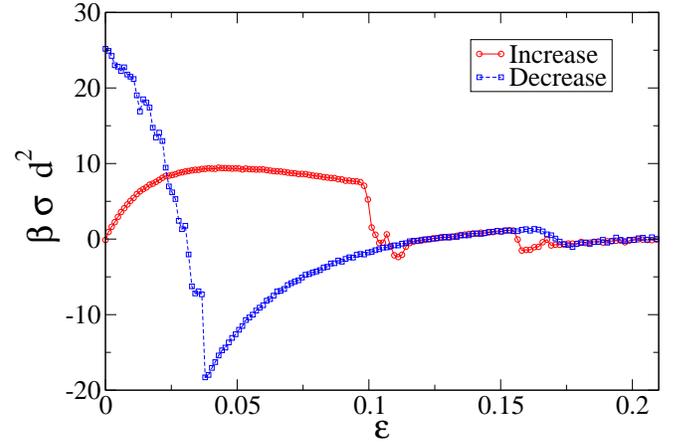}
\end{center}
\caption{ (Colour online)
A plot of the deviatoric stress $\sigma$ versus external strain 
$\epsilon$ obtained from our MC simulations of $65 \times 10$
hard disks, in presence of PBCs in both the directions, 
initially a triangular lattice at $\eta = 0.85$. Data  is obtained by 
holding at each strain value for $2\times10^4$ MCS and averaging 
over a further $3\times10^4$ MCS.  The entire cycle of increasing 
$\epsilon (\circ)$ and decreasing to zero $(\Box)$ 
using typical parameters appropriate for an atomic system,
corresponds to a real frequency of $\omega\approx 100 {\rm K\,Hz}$. 
The line through the points is guide to the eye.}
\label{stress-pbc}
\end{figure}

Reversible plasticity in the confined narrow strip is in stark contrast with 
the mechanical response of a similar strip in the absence 
of confinement. In order to show this, we study a similar narrow strip of
$65\times 10$ hard disks but now we use PBCs in {\em both} the directions. At
packing fraction $\eta=0.85$ the initial geometry ($L_x^0=65 a_x$, $L_y=10 a_y$
with $a_y=\sqrt 3 a_x/2$) contains a perfect triangular lattice. We impose
strain in a fashion similar to that described in Fig.\ref{stress}. 
The resulting stress-strain curve is shown in Fig.\ref{stress-pbc}. With
increase in strain $\e$ the system first shows a linear (Hookean)
response in the deviatoric stress $\s=\sx-\sy$, flattening out at
the onset of plastic deformation below $\e\sim 0.1$. Near $\e=0.1$ with a
solid-liquid transition (solid order parameter drops to zero with divergence
of Lindeman parameter at the same strain value) the deviatoric stress
$\s$ decreases sharply to zero in the liquid phase
obeying the Pascal's law. Unlike in the confined strip, with further increase 
in strain $\s$ does not become negative and fluctuates around zero in 
absence of wall induced density modulations. 
With decrease in strain, starting from the liquid phase,
the stress $\s$ {\em does not} trace back its path. 
Instead first a large negative stress is built up in the system as
we decrease strain up to $\e\sim0.04$. With further decrease in $\e$, the stress
$\s$ starts to increase and at $\e=0$ the system is under a huge residual
stress $\be\s d^2 =25$. The configuration of the strip at this point shows 
a solid with lattice planes rotated with respect to the initial stress-free 
lattice. This solid contains defects. Note that in the presence of hard 
confining walls, global rotation of the lattice planes cost large amounts of 
energy and would be completely suppressed. The generation of defects is also 
difficult in a confined system unless they have certain special 
characteristics which we describe later. 

\begin{figure}[t]
\begin{center}
\includegraphics[width=8.0cm]{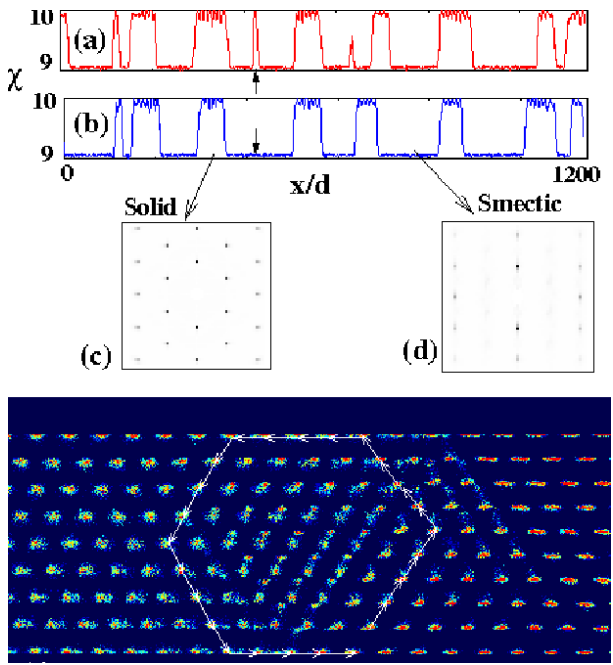}
\end{center}
\caption{(Colour online)
Plot of $\chi(x,t)$ as a function of the  channel length
$x/d$ at $\eta = 0.76$ 
after time $t=$ (a)$5\times10^5$ and (b)$2\times10^6$ MCS for 
$N = 10^3\times 10$ with $L_y=9d$. 
Note that $\chi=10$ in the solid and $\chi=9$ in the 
smectic regions. Arrows show the coalescence of two smectic bands as a function 
of time. Calculated diffraction patterns for 
the (c) $10$-layered solid  and (d) $9$-layered smectic regions.
(e) Close up view of a crystal-smectic  interface from $10^3$ superimposed 
configurations at $\eta = 0.77$. The colour 
code of the local density is the same as Fig.\ref{sld}. 
Note the misfit dislocation in the inter-facial region. A Burger's  circuit
is shown to identify the Burger's vector corresponding to this dislocation as 
$\vec b=\hat y a_y/2 + \hat x a_x/2$.
}
\label{interface}
\end{figure}

For a confined Q1D strip in the density range
$\eta_{c_2} < \eta < \eta_{c_1}$ we observe that the smectic  
order appears within narrow bands (Fig. \ref{interface}). 
Inside these bands the number of layers is less by one and the system 
in this range of $\eta$ is in a mixed phase. A plot (Fig.\ref{interface} 
(a) and (b)) of $\chi(x,t)$, where we treat $\chi$ as a space and time 
(MCS) dependent ``order parameter'' (configuration averaged number of 
layers over a window in $x$ and $t$),
shows bands in which $\chi$ is less by one compared to the 
crystalline regions. After nucleation, narrow bands coalesce to form wider
bands over very large time scales. The total size of such
bands grow as $\eta$ is decreased. Calculated diffraction 
patterns (Fig. \ref{interface} (c) and (d)) show that, locally, within a 
smectic band $S_{\bf G_1} \gg  S_{\bf G_2}$ in contrast to the solid 
region where $S_{\bf G_1} \approx S_{\bf G_2} \neq 0 $. 

Noting that $\chi = \chi^{\ast} = n_l - 1/2$ when the solid fails 
(Fig. \ref{stress} inset), it follows from Eq. \ref{strain}, the critical 
strain $\varepsilon_d^{\ast} = (4 n_l - 5)/(2 n_l - 3)(2 n_l - 2) \sim 1/n_l$. 
This is supported by our simulation data over the range $9 < n_l < 14$.
This shows that thinner strips (smaller $n_l$) are more resistant to failure.
At these strains the solid generates bands 
consisting of regions with one less particle layer. Within these 
bands adjacent local density peaks of the particles 
overlap in the $x$ direction producing a smectic. 
Within a simple density functional argument\cite{rama} it can be shown
that the spread of local density profile along $x$-axis 
$\a_x\sim 1/\sqrt{S_{\bf G_2}}$ and that along $y$-direction is 
$\a_y\sim 1/\sqrt{2 S_{\bf G_1} + S_{\bf G_2}}$~\cite{myijp}.
In the limit $S_{\bf G_2}\to 0$ (melting of solid) $\a_x$ diverges
though $\a_y$ remains finite as $S_{\bf G_1}$ remains positive definite.
Thus the resulting structure has a smectic symmetry. 

\begin{figure}[t]
\begin{center}
\includegraphics[width=8.6cm]{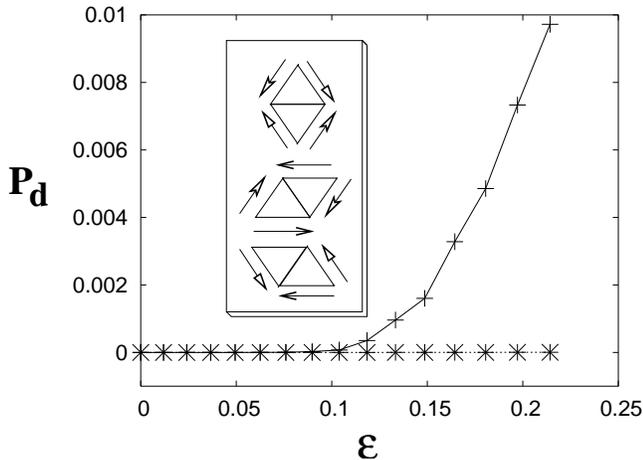}
\end{center}
\caption{Dislocation probabilities of a $65\times 10$ system are plotted as a 
function of strain starting from a triangular solid at $\eta=0.85$ and
$L_y=9d$. The corresponding bond-breaking moves are depicted in
the inset panel. Arrows show the directions of the bond-breaking moves. 
$+$ show dislocation probabilities for those Burger's vectors which have 
components perpendicular to the walls (topmost figure in the inset). On the 
other hand probabilities corresponding to the other 
two types of moves (denoted by $\ast$) remain zero.}
\label{dislo}
\end{figure}

A superposition of many particle positions 
near a solid-smectic interface [see Fig. \ref{interface}(e)] shows that: 
$(1)$~The width of the interface is large, spanning about $10 - 15$ particle 
spacings.
$(2)$~The interface between $n_l$ layered crystal and $n_l -1$ 
layered smectic contains a {\em dislocation} with Burger's 
vector in the $y$- direction which makes up for the difference in the number 
of layers.  Each band of width $s$ is therefore held in place by a 
dislocation-anti-dislocation pair (Fig. \ref{interface}). 
In analogy with classical nucleation theory\cite{pcmp,cnt}, the 
free energy $F_b$ of a single band can be written as 
\begin{equation}
 F_b = -\delta F s + E_c + \frac{1}{8\pi}b^2 K^\Delta \log \frac{s}{a_0}  
\label{becker-doring}
\end{equation}
where $K^\Delta$ is an elastic constant,
${\vec b} = \hat y a_y/2 + \hat x a_x/2$ 
is the Burger's vector,  
$\delta F$ the free energy difference between the crystal 
and the smectic per unit length and $E_c$ the core energy for 
a dislocation pair. Bands form when dislocation pairs separated by 
$s > \frac{1}{8\pi}b^2 K^\Delta/\delta F$
arise due to random fluctuations. 
To produce a dislocation pair a large energy barrier of core energy $E_c$
has to be overcome. Though even for very small strains $\varepsilon_d$ the
elastic free energy  becomes unstable the random fluctuations can not 
overcome this large energy barrier within finite time scales thereby suppressing the production of $n_l-1$ layered smectic bands up to the point of $\varepsilon_d^{\ast}$. In principle, if one could wait for truly infinite times the fluctuations {\em can} produce such dislocation pairs for any non-zero $\varepsilon_d$
though the probability for such productions $\exp(-\beta E_c)$ 
[$\beta=1/\kb T$, inverse temperature] are indeed very low. 
Using a procedure similar to 
that used in Ref.\cite{sura-hdmelt,mylif,mylif-large}, we have monitored the 
dislocation probability as a function of $\eta$ (Fig.\ref{dislo}). 
For confined hard disks, there are essentially three kinds of dislocations with
Burger's vectors parallel to the three possible bonds in a triangular solid. 
Only dislocations with Burger's vectors having components 
perpendicular to the walls, cause a change in $n_l$ and 
are therefore relevant. The dislocation formation probability is obtained
by performing a simulation where the local connectivity of bonds in the solid
is not allowed to change while an audit is performed of the number of moves 
which {\em tend} to do so. Since each possible distortion of the unit cell 
(see Fig.\ref{dislo} - inset) can be obtained by two specific sets of 
dislocations, the dislocation probabilities may be easily obtained from the 
measured probability of bond breaking moves.
Not surprisingly, the probability 
of obtaining dislocation pairs with the relevant Burger's vector
increases dramatically as $\eta \to \eta_{c_1}$ (Fig.\ref{dislo})
and artificially removing 
configurations with such dislocations suppresses the transition completely. 
Band coalescence occurs by diffusion aided dislocation ``climb'' which at  
high density implies slow kinetics. The amount of imposed strain fixes
the total amount of solid and smectic regions within the strip. The
coarsening of the smectic bands within the solid background in presence
of this conservation, leads to an even slower dynamics than non-conserving
diffusion. Therefore the size of smectic band $L(t)$ scales as 
$L\sim t^{1/3}$\cite{bray}.
Throughout the two-phase region, the crystal is in compression and the 
smectic in tension along the $y$ direction so that $\sigma$  is 
completely determined by the amount of the co-existing phases.
Also the walls ensure that orientation relationships between the two 
phases are preserved throughout. As the amount of solid or smectic in the 
system is entirely governed by the strain value $\epsilon$ 
the amount of stress $\sigma^C$ is completely determined by the value of strain
at all times regardless of deformation history. 
This explains the reversible\cite{onions} plastic deformation in 
Fig. \ref{stress}. 
   
\section{The Mean Field  Phase Diagram}
\label{phase-diagram}
In this section we obtain the phase diagram from Monte Carlo simulations
of $65\times n_l$ hard disks, $n_l$ being the number of layers contained
within the channel. We compare this with a MFT calculation.
At a given $L_y$ we start with the largest number of
possible layers of particles and squeeze $L_x$ up to the limit of overlap. Then
we run the system over $5\times 10^4$ MCS, collect data over further 
$5\times 10^4$ MCS and increase $L_x$ in steps such that it reduces the 
density by $\d\eta=0.001$ in each step. 
At a given $L_y$ and $n_l$ the value of $L_x$ that supports a
perfect triangular lattice has a density 
\bea
\eta_\triangle = \frac{\sqrt 3 \pi}{8} \frac{n_l(n_l-1)}{L_y(L_y-1)}.
\eea
In this channel geometry, at phase coexistence, the stress component 
$\sx$ of the coexisting phases becomes equal. The other component $\sy$
is always balanced by the pressure from the confining walls.
Thus we focus on the quantity $p_x=-\s_{xx}$ as $L_x$ is varied:
\bea
p_x =-\f{1}{L_y}\f{\p F}{\p L_x} = \r^2\f{\p f_N}{\p \r},~
\f{\p f_N}{\p \r} = \f{p_x}{\r^2}.
\eea
The free energy per unit volume $f_V=(N/V)(F/N) = \r f_N$. 
One can obtain the free energy  at any given density $\r$
by integrating the above differential equation starting from a known
free energy at some density $\r_0$,
\bea
f_V(\r) = \r f_N^0(\r_0) + \r \int_{\r_0}^\r \f{p_x}{\r^2} d\r~.
\label{simufeng}
\eea

We discuss below how $f_N^0(\r_0)$ for solid and modulated liquid phases are 
obtained. The free energy of the solid phase may be obtained within a simple 
analytical treatment viz. fixed neighbor free volume theory (FNFVT), which we 
outline below. More detailed treatment is available in Ref.\cite{my-htcond}.
The free volume $v_f(\eta,\chi)$ may be obtained using straight 
forward, though rather tedious, geometrical considerations and the 
free energy $f_N^0(\eta,\chi) = -\r \ln v_f(\eta,\chi)$. 
The free volume available to a particle is computed by considering 
a single disk moving in a fixed cage formed by its nearest neighbors 
which are assumed to be held in their average positions.
The free volume available to this central particle (in units of $d^2$) is 
given entirely by the lattice separations
$b=a_0(1+\ex)$ and $h=\sqrt 3 a_0 (1+\ey)/2 $ where $a_0$ is lattice parameter
of a triangular lattice at any given packing fraction $\eta$ and
$\ex = (n_l-1)/(\chi-1)$, $\ey=1/\ex$. As stated in Sec.\ref{system},
$\chi$ is obtained from channel width $L_y$ and packing fraction $\eta$.
$v_f$ is the area available to the central test particle.
Note that the effect of the confining geometry is incorporated in
the lattice separations $b$, $h$.
The FNFVT free energy has minima at all $\chi= n_l$. For half 
integral values of $\chi$ the homogeneous crystal is locally unstable. Although
FNFVT fails also at these points, this is irrelevant as the system 
goes through a phase transition before such strains are realized. 
In the high density triangular solid phase, we know $f_N^0(\r_0=\r_\triangle)$,
exactly, from the fixed neighbor free volume theory (FNFVT). It is 
interesting to note that, apart from at $\r_\triangle$, this FNFVT can
estimate the solid free energy quite accurately at small strains $\ex <4\%$
around $\r_\triangle$. To obtain the 
solid branch of free energy $f^S_V(\r)$ from simulations, we integrate the 
$p_x - \r$ curve starting from the FNFVT estimate at 
$\r=\r_0\equiv \r_\triangle$. 

For the confined fluid
phase we use a phenomenological free energy~\cite{santos} of hard disk fluid
added with a simple surface tension term due to the scaled particle 
theory~\cite{spt}. We outline this in the following. 
For the bulk (unconfined) liquid phase one can obtain $f_N^0(\r=\r_l)$ up to a 
fairly large volume fraction $\eta_l\sim 0.81$ where the liquid reaches its 
metastable limit in bulk, using the phenomenological expression\cite{santos},
\bea
\be f_N^0(\r_l) &=& \f{(2\eta_c-1)\ln(1-\f{2\eta_c-1}{\eta_c}\eta_l) 
- \ln(1-\f{\eta_l}{\eta_c})}{2(1-\eta_c)}\nn\\
 &+& \ln\r_l -1  
\eea
where $\eta_c$ is the close packed volume and depends on $L_y$. We
have checked from the simulations that this equation of state
shows good agreement with the simulated $p_x-\eta$ diagram in a confined channel
up to a density of $\eta=0.65$. Above this density, $p_x$ falls below
and $p_y$ goes above this equation of state estimate.
Integrating $p_x$ from $\eta_0=\eta_l\equiv 0.6$ using the above 
mentioned form of $f_N^0(\r_l)$ we find the bulk part of the liquid branch of 
free energy $f^L_V(\r)$. In order to incorporate the effect of confinement
approximately, we add  a scaled particle 
theory (SPT)~\cite{spt} surface energy contribution to the bulk free energy. 
Within SPT, the interfacial tension (interfacial energy per unit length) 
of a hard disk fluid in contact with a hard wall can be calculated 
as $\g = \bar\g + p_{spt} d/2$ \cite{sptlowen} where 
$\bar\g=(\r d/2\be)[1/(1-\eta)-1/(1-\eta)^2]$ and the SPT equation of state
is $\beta p_{spt}/\r =1/(1-\eta^2)$~\cite{spt}. Thus the SPT surface energy
per unit volume $S_T=\g/L_y=\r d/2\beta(1-\eta)L_y$ is added with the bulk
contribution $f^L_V(\r)$ to account for the confinement.

\begin{figure}[t]
\begin{center}
\includegraphics[width=8.6cm]{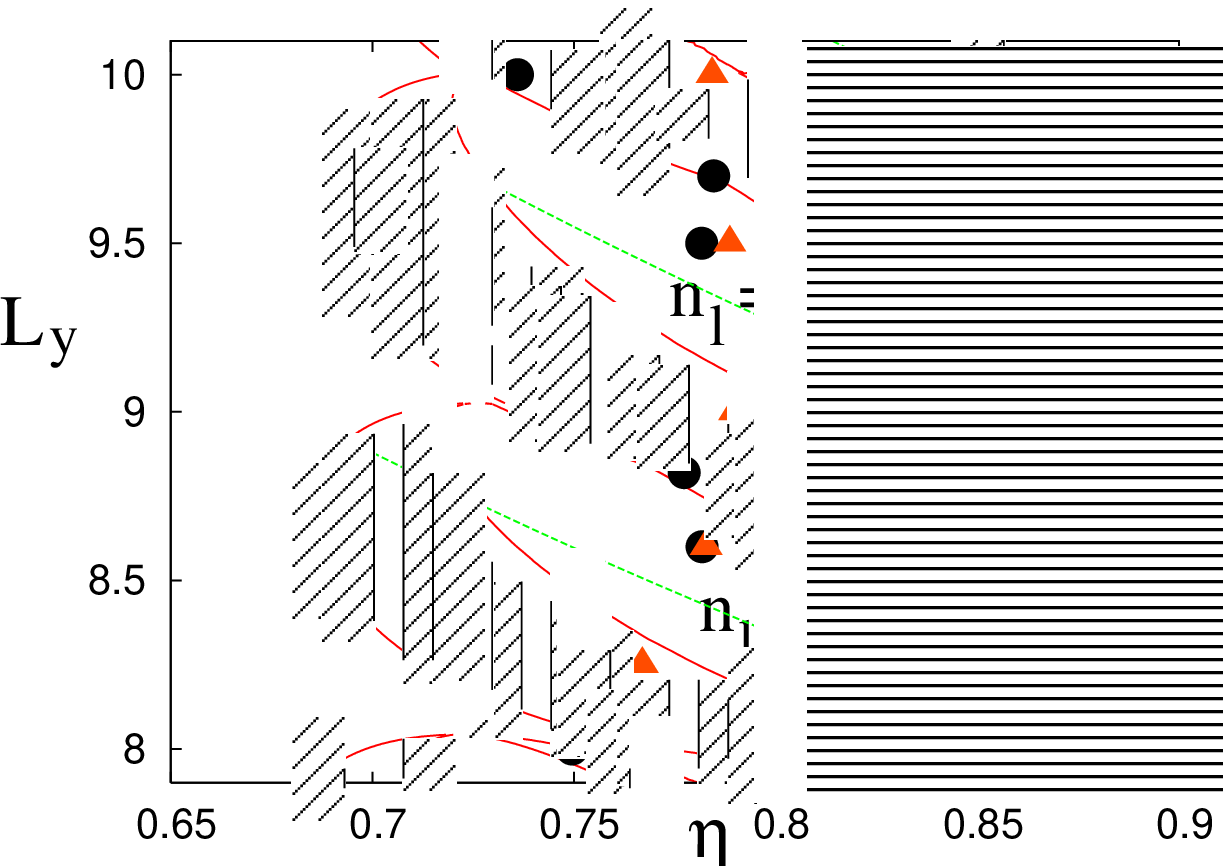}
\end{center}
\caption{(Colour online) Phase diagram.
The continuous lines denote the phase boundaries obtained from 
mean field theory. 
The white,  vertically shaded and horizontally shaded regions indicate
single phase, two-phase coexistence and forbidden regions, respectively.
The single phase regions of low densities are liquid and high densities 
are solid phases.  On the green (dashed) lines  $n_l=9,10,11$ the system 
is commensurate with the channel width such that exactly integral number 
of layers are contained.  Points denoted by filled $\circ$,  
filled $\triangle$, and filled $\Box$ are obtained from simulations.
Regions with densities less than the filled $\circ$ are modulated liquid and 
above filled $\triangle$ are solid. In the intervening region, we have a 
solid-smectic coexistence. The filled $\Box$ mark the  onset of 
buckling instability. Smectic order increases smoothly and continuously as 
density increases. The liquid within the coexistence region is highly 
modulated.}
\label{phdia}
\end{figure}

At thermodynamic equilibrium, the chemical potential
$\mu = -(\p f_V/\p \r)_{V,T}$ and pressure for the coexisting phases
must be same. Thus from a
plot of free energy vs. density obtaining common tangent to solid and liquid 
branches of free energies means finding coexisting densities of solid and 
liquid having the same chemical potential. 
Once the two branches of free energy are obtained 
in the above mentioned way, we fit each branch using  polynomials 
up to quadratic order in 
density. The common tangent must share the same slope and intercept. 
These two conditions uniquely determine 
the coexisting densities. The onset of buckling instability can be found
from the slope of the $p_x-\eta$ curve at high densities -- a negative 
slope being the signature of the transition. The estimates of the coexisting 
densities and onset of the buckling instabilities in our simulations are thus 
obtained and shown using filled symbols in the phase diagram (Fig.\ref{phdia}).

To obtain the phase diagram from MFT (lines in Fig.\ref{phdia})
we use the above mentioned
phenomenological free energy of hard disk fluid\cite{santos} together with
the SPT surface tension term. For the solid we use the FNFVT free energy for 
all densities. We take care to minimize the free energy of the solid with 
respect to choices of $n_l$ first. Then we find out the minimum of the solid
and liquid free energies at all densities. The coexisting densities are 
obtained from common tangent constructions for this minimum free energy.
Our MFT phase diagram so obtained is shown along with the simulation 
results in the $\eta-L_y$ plane (Fig.\ref{phdia}). 

In Fig.\ref{phdia} points are obtained from simulations and the continuous 
curves from theory. The regions with
densities lower than the points denoted by filled circles are single phase
modulated liquid. Whereas the regions with densities larger than 
the points denoted by filled triangles indicate the single phase solid.
The modulations in the liquid increase with increasing density and at high 
enough $\rho$ the structure factor shows only two peaks typical of the 
smectic phase. This transition is a smooth crossover. 
All the regions in between the filled circles and filled triangles
correspond to solid-smectic coexistence.  The filled
squares denote the onset of buckling at high densities.
The high density regions shaded by horizontal lines are physically 
inaccessible due to the hard disk overlap. 
The MFT prediction for the solid-liquid coexistence is shown 
by the regions between the continuous (red) lines (shaded vertically). 
The unshaded white regions bounded by red lines and labeled by the number 
of layers denote the solid phase. All other unshaded regions denote liquid 
phase. For a constant channel width, a discontinuous transition from liquid to 
solid via phase coexistence occurs with increasing density. However, the  
MFT predicts that the solid remelts at further higher densities.  
Notice that the simulated points for the onset of buckling lie very close 
to this remelting curve. Since the MFT, as an input, has 
only two possibilities of solid and fluid phases, the high density remelting
line as obtained from the MFT may be interpreted as the onset of
instability (buckling) in the high density solid. 
The MFT prediction of the solid-fluid coexistence
region shows qualitative agreement with simulation results for
solid-smectic coexistence. The area of phase digram corresponding to the
solid phase as obtained from simulation
is  smaller than that is predicted by the MFT calculation. This may be due
to the inability of MFT to capture the effect of fluctuations. 
From the simulated phase diagram
it is clear that if one fixes the density and channel width in a solid phase 
and then increases the channel width keeping the density fixed, one finds
a series of phase transitions from a solid to a smectic to another solid
having a larger number of layers. These re-entrant transitions are due to the
oscillatory commensurability ratio $\chi$ that one encounters on changing the 
channel width. This is purely a finite size effect due to the confinement.
It is important to note that for a bulk hard disk system, solid phase
is stable at $\eta>0.719$,
whereas, even for a commensurate confined strip of hard disks, the 
solid phase is stable only above a density $\eta=0.75$. With increase in
incommensuration the density of the onset of solid phase increases further. 
This means that confinement of the Q1D strip by planar walls has an overall
{\em destabilizing} effect on the solid phase.  

The phase diagram calculated in this section is a MFT phase diagram 
where the effects of long wavelength (and long time scale) fluctuations are 
ignored.  For long enough Q1D strips fluctuations in the displacement
variable should increase linearly 
destroying all possible spontaneous order and leading to a single disordered
fluid phase which is layered in response to the externally imposed hard wall 
potential. However, even in that limit the layering transitions from lower 
to higher number of layers as a function of increasing channel width, 
survives\cite{degennes}.

Our simple theory therefore predicts a discontinuous solid-fluid transition 
via a coexistence region with change in density or channel width. However, 
details like the density modulation, effects of asymmetry in density profile, 
vanishing displacement modes at the walls and most importantly nucleation and 
dynamics of misfit dislocations crucial to generate the smectic band mediated 
failure observed in simulations are beyond the scope of this theory. Also, the 
effect of kinetic constraints which stabilize the solid phase well inside the 
two phase coexistence region is not captured in this 
approach. We believe, nevertheless, that this equilibrium calculation 
may be used as a basis for computations of reaction rates for addressing 
dynamical questions in the future. 

Before we end this section, it  is useful to compare our results with
similar calculations in higher dimensions viz hard {\em spheres} confined
within hard parallel plates forming a quasi two dimensional 
film\cite{buckled-1,fortini,buckled-3}. Commensuration effects produce
non-monotonic convergence of the phase boundaries to that of the bulk
system. The appearence of novel phases e.g. the buckled solid not observed 
in the bulk is also a feature showed by all these systems; on the other
hand these quasi two dimensional systems show two kinds of solids -- 
triangular and square -- and no `smectic like' phase. The effect of 
fluctuations also should be much less in higher dimensions and one expects 
the observed phases to have less anomalous properties.

\section{Discussion and Conclusion}
\label{conclusion}
\begin{figure}[t]
\begin{center}
\includegraphics[width=8.0cm]{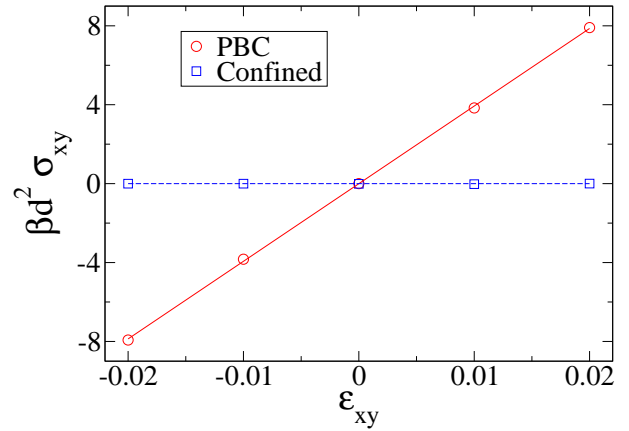}
\end{center}
\caption{(Colour online)
Shear stress vs. shear strain at $\eta=0.85$. A system of 
$40\times 10$  hard disks simulated with periodic boundary conditions 
and $L_y$ commensurate with ten layered triangular solid gives
a shear modulus $\mu=398\pm 4$. On the other 
hand when the same system of $40\times 10$ 
hard disks is confined within a commensurate channel,
that fits $10$-layers of lattice planes, the shear modulus drops drastically 
to $\mu=0$.}
\label{shear}
\end{figure}

One of the key definitions of a solid states that a solid, as opposed to 
a liquid, is a substance which can retain its shape due to its nonzero 
shear modulus. Going by this definition, a Q1D solid strip even of finite
length confined within planar, structureless walls is not a solid despite 
its rather striking triangular crystalline order. Indeed, the shear modulus 
of the confined solid at $\eta=0.85$ is zero, though the corresponding system
with PBC show finite nonzero shear modulus (See Fig.\ref{shear}).
This is a curious result and is generally true for all values of 
$4 < n_l < 25$ and $\eta$ investigated by us.
Confinement induces strong layering which effectively decouples the
solid layers close to the walls allowing them to slide past each other
rather like reentrant laser induced melting\cite{frey-lif-prl,lif-hd,mylif}.
This immediately shows that the only thermodynamically stable phase in
confined Q1D channel is a modulated liquid, the density modulation 
coming from an explicit breaking of the translational symmetry by the 
confinement.

\begin{figure}[t]
\begin{center}
\includegraphics[width=8.0cm]{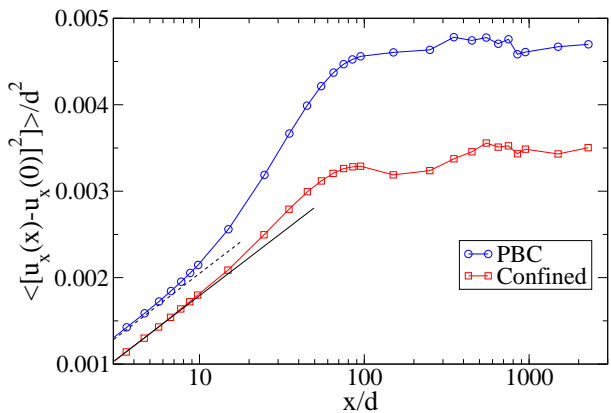}
\end{center}
\caption{(Colour online)
Flucuation $<({u^x}(x) - {u^x}(0))^2>/d^2$ on a lattice plane (line)
for $5000\times 10$ commensurate triangular solid ($\eta=0.85$) with PBC 
in both directions is compared with that in presence of hard channel 
confinement.
Averages are done over $100$ configurations.
$\circ$ denotes the data obtained in presence of PBC in both directions and
$\Box$ denotes the data obtained for confined system. In both cases 
the initial logarithmic growth (tracked by dashed and solid straight
lines respectively) crosses over to a faster linear increase before 
saturating due to finite size effects. Confinement apparently reduces the
amount of fluctuations without afecting the general behavior.}
\label{uxux-85}
\end{figure}

To understand the nature and amount of fluctuations in the confined Q1D
solid we calculate the auto-correlation of the displacement along the channel, 
$B(x)=\la({u^x}(x) - {u^x}(0))^2\ra$ for a layer of particles near a boundary. 
The nature of the {\em equilibrium} displacement correlations ultimately 
determines the decay of the peak amplitudes of the structure factor 
and the value of the equilibrium elastic moduli~\cite{pcmp}. 
It is known from harmonic theories, in one dimension 
$B(x) \sim x$ and in two dimensions $B(x)\sim \ln (x)$. 
In the Q1D system it is expected that 
for small distances, displacement fluctuations will grow logarithmically 
with distance which will crossover to a linear growth at large 
distances with a crossover length $x_c\sim L_y \ln L_y$~\cite{ricci,ricci2}.    
We calculate this correlation for a system of $5000\times 10$ particles 
averaged over $10,~30,~50,~100$ 
configurations each separated by $10^3$ 
MCS and equilibrated over $10^5$ MCS at $\eta=0.85$.
We compare the results obtained for a strip with confinement and a strip with
PBC's in both directions, taking care that in each case the channel width is
commensurate with the inter-lattice plane separation.
With an increase in the number of configurations 
over which the averages are done, fluctuations reduce and converge to a 
small number. This happens for both the cases we study.
It is interesting to notice a logarithmic to linear cross-over of the
fluctuations near $x=10d$ for both the cases. Since the harmonic 
theory\cite{ricci2} ignores effects due to boundaries, one can conclude
that this general behavior is quite robust. 
At large distances, 
displacement correlations are expected to saturate due to finite size effects. 
The magnitude of fluctuation at saturation for the confined system 
($\sim 0.0035$) is noticably lower than that in presence of PBC 
($\sim 0.0045$) (Fig.\ref{uxux-85}). Thus the introduction of commensurate 
confinement can have a {\em stabilizing} effect of reducing the overall 
amount of fluctuations present in a Q1D strip with PBC.

We have separately calculated the elastic modulus of a $40 \times 10$ confined 
hard disk solid at $\eta =0.85$ under elongation (Young's modulus) in the $x$ 
and the $y$ directions. For an isotropic triangular crystal in 2D these should 
be identical. We, however, obtain the values $1361$ and $1503$ (in units of 
$k_B T/d^2$ and within an error of $3\%$) respectively for
the two cases. The Young modulus for the longitudinal elongation is smaller 
than that in the direction transverse to the confinement and both these values 
are larger than the Young modulus of the system under PBC ($Y=1350$).
This corroborates the fact that the non-hydrodynamic component 
of the stress $\s_{xx}-\s_{yy}$ is non-zero even for vanishingly small 
strains as shown in Fig.\ref{stress}. Therefore even if we choose to 
regard this Q1D solid as a liquid, it is quite anomalous since it 
trivially violates Pascal's law which states that the stress tensor in a 
liquid is always proportional to the identity. This is  because
of the explicit breaking of translational symmetry by the confinement.
Lastly, commensurability seems to affect strongly the nature and magnitude 
of the displacement fluctuations which increase dramatically as the system 
is made incommensurate.  

As $L_x\approx L_y \to \infty$ all such anomalous behavior is 
expected to be localized in  a region close to the walls such that
in the central region a bulk 2d solid is recovered. This crossover
is continuous, though oscillatory with commensurabilities playing
an important role, and extremely slow ($\sim 1/L_y$). It is 
therefore difficult to observe in simulations.

What impact, if any, do our results have for realistic systems?  
With recent advances in the field of nano science and 
technology\cite{nanostuff-1,nanostuff-2} new possibilities of building
machines made of small assemblage of atoms and molecules are
emerging. This requires a knowledge of the structures and mechanical 
properties of systems up to atomic scales. A priori, at such small scales,
there is no reason for macroscopic continuum elasticity 
theory, to be valid\cite{micrela}. Our results corroborate such expectations.
We have shown that small systems often show entirely new behavior if 
constraints are imposed leading to confinement in one or more directions. 
We have also speculated on applications of reversible failure as accurate 
strain transducers or strain induced electrical or thermal switching 
devices~\cite{my-econd,my-htcond}. We believe that many of our results may 
have applications in tribology\cite{ayappa} in low dimensional systems. The 
effect of corrugations of the walls on the properties of the confined system 
is an interesting direction of future study. The destruction of long ranged 
solid like order should be observable in nano wires and tubes and may lead to 
fluctuations in transport quantities \cite{akr}. 

\vskip .2cm
\section{acknowledgment}
The authors thank M. Rao, V. B. Shenoy, A. Datta, A. Dhar, A. Chaudhuri
and A. Ricci for useful discussions. Support from SFB-TR6 program on 
``Colloids in external fields" and the Unit for Nano Science and 
Technology, SNBNCBS is gratefully acknowledged.

\end{document}